# Willis metamaterial on a structured beam

Yongquan Liu[1,2], Zixian Liang[3], Jian Zhu[2], Lingbo Xia[2,5], Olivier Mondain-Monval[4], Thomas Brunet[6], Andrea Alù[5*], Jensen Li[1,2*]

[1] Department of Physics, The Hong Kong University of Science and Technology, Clear Water Bay, Hong Kong, China
[2] School of Physics and Astronomy, University of Birmingham, Birmingham B15 2TT, United Kingdom
[3] College of Electronic Science and Technology, Shenzhen University, Shenzhen 518060, China
[4] University of Bordeaux, CNRS, UPR 8641, CRPP, 33600 Pessac, France
[5] Department of Electrical and Computer Engineering, The University of Texas at Austin, Austin, TX 78712, USA.
[6] University of Bordeaux, CNRS, UMR 5295, I2M-APy, 33405 Talence, France

*Corresponding author: aalu@gc.cuny.edu; jensenli@ust.hk

**Abstract**

Bianisotropy is common in electromagnetics whenever a cross-coupling between electric and magnetic responses exists. However, the analogous concept for elastic waves in solids, termed as Willis coupling, is more challenging to observe. It requires coupling between stress and velocity or momentum and strain fields, which is difficult to induce in non-negligible levels, even when using metamaterial structures. Here, we report the experimental realization of a Willis metamaterial for flexural waves. Based on a cantilever bending resonance, we demonstrate asymmetric reflection amplitudes and phases due to Willis coupling. We also show that, by introducing loss in the metamaterial, the asymmetric amplitudes can be controlled and can be used to approach an exceptional point of the non-Hermitian system, at which unidirectional zero reflection occurs. The present work extends conventional propagation theory in plates and beams to include Willis coupling, and provides new avenues to tailor flexural waves using artificial structures.

Metamaterials, constructed with artificially designed microstructures, have been employed and developed in electromagnetism[1,2], acoustics[3-7], thermodynamics[8,9] and mechanics[10-12] to give unique properties beyond those provided by natural and composite materials. Such a concept has recently been extended to elastic waves in solids[13-17]. Compared to acoustic waves, the additional degrees of freedom in polarizations states require more sophisticated dispersion engineering[18-20] and also lead to non-trivial mode-matching at interfaces. Metamaterials are found particularly useful as a tool to explore the physics behind this complexity. For example, for in-plane waves, a trans-modal Fabry-Pérot condition is found necessary for maximum mode conversion between longitudinal and shear modes[21]. For flexural waves[22,23], evanescent modes can be used to modify surface impedance for obtaining higher transmission than structures in acoustics[24].

Peculiarly, classical elastic wave equations, i.e. the Hooke's law together with the Newton's second law, are not form-invariant under a coordinate transformation[25], suggesting that we are currently exploring an unnecessarily limited palette of material properties. For example, an introduction of rotation modulus in Hooke's law is found necessary to describe biological composites such as wet bones[26]. This is called a Cosserat solid and it has been recently constructed using metamaterial approach[27]. Another attempt is the proposal to realize Willis media[28]. These media introduce new constitutive terms not only in Hooke's law but also in Newton's second law[29-33]. These terms, which introduce coupling between stress and velocity and between momentum and strain, are typically small perturbations and difficult to realize. Currently, a similar modification to both equations was experimentally introduced in airbone acoustics[34-36], and a strategy to induce strong Willis coupling in suitably designed acoustic metamaterials was theoretically introduced[37]. The acoustic wave equations, when Willis coupling is considered, can be written in analogy to the electromagnetic scenario, and the additional constitutive terms correspond to bianisotropy in electromagnetism[38-41]. These developments indicate that Willis media for elastic waves in solid may become practical through the notion of metamaterials, although the exact microstructural design has yet to be determined.

In this work, we consider the situation when a Willis medium is used to construct a plate or a beam for flexural wave propagation. Such a reduced version from three to two dimensions facilitates analysis and an intuitive understanding of Willis coupling. Plates and beams are actually common in a wide range of length scales from building structures to micromechanical systems. Their theories can be traced back to the 1950s to 1980s in a series of works from Kirchhoff-Love plate theory to extensions with rotary and shear deformations, and from isotropic to anisotropic plates[42,43]. However, further extension to include Willis coupling to plate theory is deemed necessary to open new directions in this field of research. The special responses offered by Willis coupling, including asymmetric reflection amplitudes and exceptional points, open an unexplored territory for ultrasonic nondestructive evaluation, seismology and micro-sensor technology[44,45].

**Willis plate theory**

We start from a Willis medium, with constitutive relation in its most general form[31]:

$$\sigma_{ij} = C_{ijkl}\varepsilon_{kl} - \omega^2 S_{ijk}u_k,$$
$$p_j = -i\omega S_{klj}\varepsilon_{kl} - i\omega\rho_{ij}u_i, \qquad (1)$$

where $C$ and $\rho$ are the stiffness and density tensors, $\omega$ is the radial frequency and $i, j, k$ are indices iterating the spatial coordinates. The additional $S$ term in Hooke's law, Willis coupling coefficient, couples stress $\sigma$ to displacement $u$ and the same term (due to reciprocity) in Newton's second law couples momentum $p$ to strain $\varepsilon$. Suppose that we now construct a plate ($z = -h/2$ to $h/2$) from this medium. By integrating Eq. (1) along $z$ (zeroth moment), we have

$$Q_x = 2\mu h \varepsilon_{xz} - \omega^2 h \tau u_z,$$
$$P_z = -2i\omega h \tau \varepsilon_{xz} - i\omega \rho_z h u_z. \qquad (2)$$

We assume wave propagation along the x-direction ($\partial_y \to 0$ for simplicity). Eq. (2) relates the shear force $Q_x = \int \sigma_{xz} dz$ and the total momentum $P_z = \int p_z\, dz$ to $u_z$ and $\varepsilon_{xz}$. $\mu$ and $\rho_z$ are recognized as the effective shear modulus and density along $z$ for the plate. The additional constant $\tau$ (it is actually $S_{xzz}$ in the bulk) is the Willis coupling term. For plate theory, we actually need to integrate up to the first moment ($\int z dz$) Eq. (1) and the equation of motion, yielding

$$\partial_x^2(D\partial_z\varepsilon_{xx}) + \frac{h^3\omega^2}{12}\partial_x(\rho_x\partial_z u_x) = \partial_x Q_x = -i\omega P_z \qquad (3)$$

which has the same form in Mindlin plate theory[42], but now with Willis coupling terms in $Q_x$ and $P_z$ to describe propagation of $u_x$ and $u_z$. $D$ is the bending stiffness defined for either a plate or a beam[46]. The second term involving $\rho_x$, density along $x$, refers to rotary motion, which can be neglected when the plate is sufficiently thin. Eq. (2) and (3) describes wave propagation along $x$-direction for a Willis plate or beam although we have not yet designed the mechanism to generate the required Willis coupling. We also note that there can be possibly a higher-order Willis coupling term in the first moment (which will appear inside the two brackets in Eq. (3)), but its effect is negligible for subwavelength inclusions, see Supplementary Notes 1 and 2 for more details).

**Metamaterial design with Willis coupling**

As the Willis coupling term $\tau$ comes from $S_{xzz}$ of the bulk, a necessary condition to have non-zero $\tau$ is broken mirror symmetry in the x-direction. Figure 1a shows the unit cell of the metamaterial plate designed by perforating slots in a background acrylic plate. The inner disk is connected to the matrix by two thin ribs, either along the y-direction (as a reference case without Willis coupling in the next section), or both connected to the back interface (as in the case with Willis coupling). The various dimensions of the structure are listed in the caption. In the latter case, there is a cantilever bending resonance for the central disk at around 14.5kHz (see Supplementary Note 4), designed to implement a strong Willis coupling: when there is a constant force $F_z$ applied along the $z$-direction on both the front and back interfaces, a typical simulated displacement profile $u_z$ around the resonating frequency is shown as the color map. The back interface, being dragged by the disk, has a smaller magnitude of displacement than the front interface. This asymmetry gives rise to non-zero Willis coupling between $P_z$ (equivalent to the $F_z$) and $\varepsilon_{xz}$ in Eq. (2).

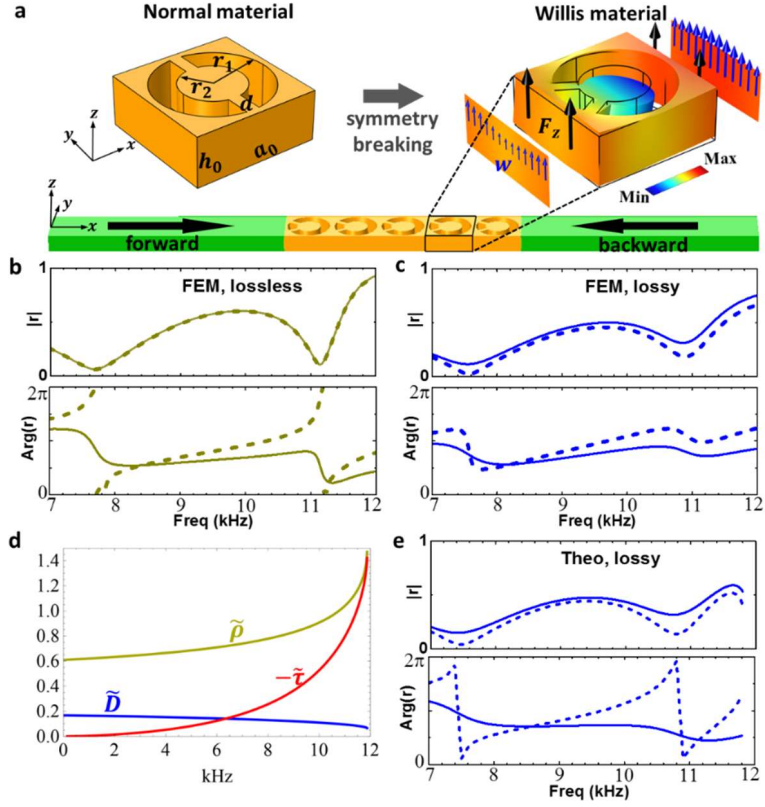

**Figure 1 | Symmetry breaking in reflection and Willis material.** **a,** Normal material is changed into Willis material by breaking mirror symmetry. For this Willis material, $a_0 = 5.0\ mm$, $h_0 = 2.0\ mm$, $r_1 = 0.45a_0$, $r_2 = 0.25a_0$, and $d = 0.5mm$ with orientation angles of two thin ribs 150 and 210 degrees, respectively. The displacement field $u_z$ is plotted by loading force $F_z$ (shown as black arrows). The field of $u_z$ on the two end surfaces are also presented by blue arrows. The Willis material is assembled as a metamaterial layer in a thin beam. Forward/backward plane wave is incident into the left/right side of the beam. **b,** Amplitudes $|r|$ and phases $\mathrm{Arg}(r)$ of reflected waves from both sides for lossless material (density $\rho_0 = 1190 kg/m^3$, Young's modulus $E_0 = 2.95 GPa$ and Poisson ratio $\nu_0 = 0.29$). **c,** Amplitudes $|r|$ and phases $\mathrm{Arg}(r)$ of reflected waves for lossy material (the Young's modulus turns into $E_0 = 2.95(1 - 0.05i)GPa$). **d,** Simulated results of effective parameters. **e,** Theoretical values of reflection amplitudes $|r|$ and phases $\mathrm{Arg}(r)$ for lossy material. In **b**, **c** and **e**, the solid/dashed lines corresponds to the case of forward/backward incident waves.

This asymmetry of displacement field is connected to asymmetric reflection. Suppose that 5 unit cells are cascaded and embedded in the background plate. The forward and backward reflection amplitudes, $r_f$ and $r_b$, are obtained by full-wave simulations from 7 to 12 kHz (corresponding to wavelengths from 28.1 to 21.2 mm) and are plotted in Fig. 1b. As the metamaterial is reciprocal, we have equal transmission $t$ in both directions and hence the same magnitude of reflection amplitude if the material is lossless: $|r_f|^2 = 1 - |t|^2 = |r_b|^2$. The asymmetry of reflection only shows up in the phase and becomes most prominent at 7.6 and 11.0 kHz, in which Fabry-Pérot (FP) resonances for elastic waves[47] occur as the reflection amplitude dips. If we add an material loss (by adding 5% of real part of Young's modulus to the imaginary part), the magnitude of the reflection amplitudes become asymmetric as well, as shown in Fig. 1c.

The asymmetry in reflection can be understood in terms of the Willis coupling parameter. Frist, we define an effective medium model of our structure from the eigenmode profiles (see Supplementary Note 2). The normalized bending stiffness $\widetilde{D} = D/D_0$, normalized density $\tilde{\rho} = \rho/\rho_0$, and normalized Willis coupling parameter $\tilde{\tau} = \frac{\tau}{\mu} \frac{\omega^2}{k_0}$ are plotted in Fig. 1d. These are the minimal set of dimensionless parameters for our case to describe the metamaterial with an effective medium model. $D_0$, $\rho_0$ and $k_0$ are the corresponding bending stiffness, density and wave number for the background plate. Then, the transfer matrix can be formulated numerically from the effective medium with thickness of 5 unit cells and the reflection amplitudes (for the lossy case) are plotted in Fig. 1e, showing similar behavior to full-wave simulation. In the long wavelength limit, the transfer matrix can be simplified analytically, giving rise to a difference of reflection amplitudes $r_f - r_b \approx -\tilde{\tau} k_0 l e^{-ik_0 l}$, where $l$ is the total thickness of the metamaterial (see Supplementary Note 3). This provides a qualitative understanding of the contribution of the Willis coupling to asymmetric reflection. When we turn off $\tilde{\tau}$ by setting it to zero, the reflection amplitudes become the same for both directions as expected (results not shown).

**Experimental verification of Willis properties**

A series of experimental studies have been carried out to demonstrate Willis coupling and the resultant asymmetric reflection. Fig. 2a shows the photographs of two experimental samples, with geometric dimensions listed in the caption. The top one shows our designed Willis material while the bottom one shows a symmetric metamaterial. A number of unit cells for either structure are 3D printed on an acrylic beam of 2 mm in thickness. Flexural waves, impinging from the left (forward) or the right (backward) directions, are then excited by piezo transducers while the out-of-plane displacement at various positions are mapped by a scanning laser vibrometer (Polytec PSV-400). The reflection and transmission amplitudes in both directions are extracted from the measured fields directly (see Method section for details).

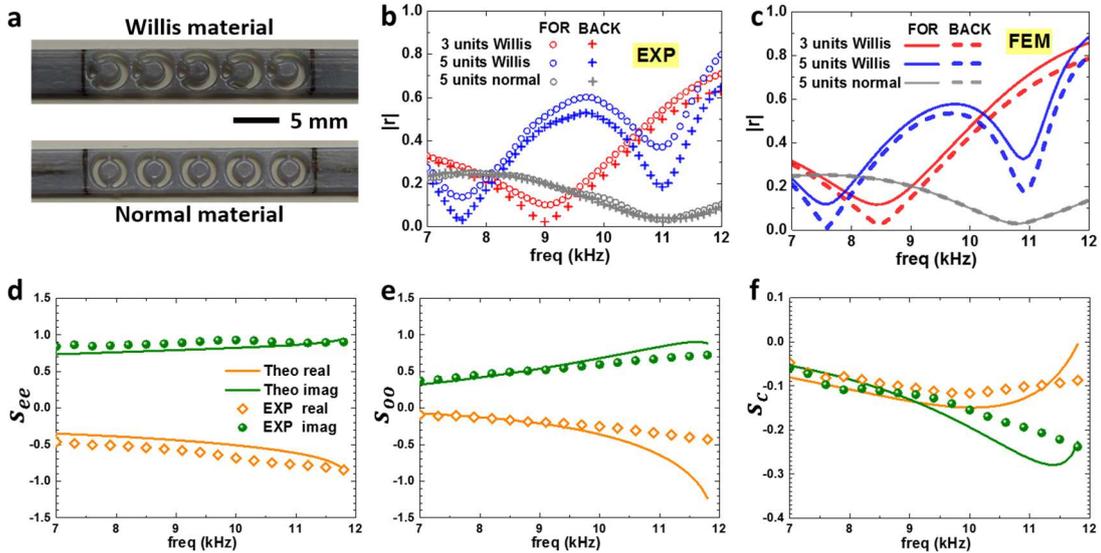

**Figure 2 | Experimental testing of asymmetric reflection and extraction of Willis coupling. a**, Photograph of normal and Willis material structures. **b and c**, The experimentally tested and simulated reflected amplitude of different cases, respectively. For the normal material, $r_1 = 0.4a_0$, $d = 0.4mm$ with orientation angles of two thin ribs $\pm 90$ degrees are set. Other geometric and material parmeters are same to the Willis cases. **d-f,** The theoretical (solid lines) and

experimental (symbols) values of the even scattering coefficient $s_{ee}$, the odd scattering coefficient $s_{oo}$ and the cross-coupling scattering coefficient $s_c$ of the Willis metamaterial.

Fig. 2b shows the reflection amplitudes given by the Willis material, shown by the blue and red symbols for 5 and 3 unit cells respectively. The reflection amplitude ($|r_b|$) in the backward direction is smaller than the one ($|r_f|$) in the forward direction. Such an asymmetry is most prominent near the FP resonances: the reflection dips at 7.7 kHz and 11.0 kHz for the 5-unit cells sample, where the difference of reflection amplitude $\Delta|r| = |r_f| - |r_b|$ reaches to local maximums 0.10 and 0.18, and at 9.0kHz for the 3-unit cells sample, where the difference reaches to 0.08. This indicates that our system actually exhibits certain amount of material loss (as assumed in last section). The multiple reflections near the FP condition within the unit cells forces the waves to interact more strongly with the bending cantilever resonance, and hence the asymmetry in amplitude becomes larger. On the contrary, the grey symbols in the same figure show the forward and backward reflection amplitude for a similar structure (but symmetric), in which no asymmetric reflection occurs. The observed asymmetry in reflection also agrees well with full-wave simulation results, shown in Fig. 2c. For example, the Willis material with 3 unit cells, exhibits the same FP dip at 8.6 kHz with quantitative agreement $\Delta|r| = 0.09$ to the experiment.

As we have measured the complex transmission and reflection coefficients for excitation from both sides, we can reveal Willis coupling directly by plotting the even and odd scattering coefficients. Similar to the case of electromagnetic metamaterials[48,49], materials without bianisotropy only convert even (odd) inputs to even (odd) scattered waves (termed as $s_{ee}$ and $s_{oo}$). The Willis coupling here contributes a cross-coupling between even input and odd output or equivalently between odd input and even output (termed as $s_c$, it represents $s_{eo} = -s_{oe}$ with the same amplitude but minus sign from reciprocity[30,31]). Figs. 2d to 2f show the even scattering coefficient $s_{ee}$, the odd scattering coefficient $s_{oo}$ and the cross-coupling scattering coefficient $s_c$. First, the scattering matrix (parameters $t$, $r_f$ and $r_b$) can be either measured from experiment for a sample with only one unit cell or obtained from theory: the unit cell scattering matrix resultant from the effective medium in last section. Then, the even and odd scattering coefficients can be obtained by linearly combining the matrix elements (see Supplementary Note 5). In our case, a non-zero cross-coupling scattering coefficient $s_c$ (approximated by $-0.1 + (0.24 - 0.04f)i$ with $f$ being the frequency in kHz), directly reveals the effective "bianisotropy" given by Willis materials. Moreover, due to symmetry breaking, the bending cantilever resonance can now be excited by both even and odd inputs, giving rise to $s_{oo}$ and $s_{ee}$. In contrary, $s_c$ becomes zero if a mirror symmetric structure is used for testing.

**Tunable asymmetry by varying the loss coefficient**

Next, we show that the degree of asymmetry and also the frequencies (the FP dips) where the asymmetry occurs can be easily tuned by loading additional materials with loss to the Willis metamaterial. Here, soft porous silicon rubber synthesized by thermal polymerization[50] is pasted onto the upper and lower surfaces of the metamaterial layer (with 5 unit cells), as schematically shown in Fig. 3a. Figure 3b shows the experimental reflection amplitudes for metamaterials with soft porous rubbers (with thickness of 2mm and porosity of 12%). Apart from tuning the loss, the pasted porous rubber largely changes the band structure and eigenmodes of the metamaterial, which leads to a different set of effective material parameters. As a result, a FP dip is observed at 10.24 kHz. Furthermore, when an appropriate amount of loss is added to the metamaterial, the reflection dip actually approaches to zero (for one direction), yielding the so-called unidirectional zero reflection (UZR), see Supplementary Note 6 for numerical results. For the case with two porous rubber sheets loaded, the reflection dip in the forward direction attains a value of 0.004, very close to zero, while the reflection amplitude in the other direction is around 0.128. The quality of the UZR can also be visualized by examining the eigenvalues of the scattering matrix $S$ in the complex plane. For a non-Hermitian system, like the one at hand, the eigenvalue degeneracy of the scattering matrix is called exceptional point of the system and a UZR occurs exactly at that point [49]. As shown in Fig. 3c, trajectories against frequency for both eigenvalues of $S$ first move towards each other, approaching the exceptional point at 10.24 kHz and then split again with increasing frequency. If we can tune the loss in a finer degree than the current experiment, it is possible to further improve the quality by bringing the two trajectories even closer to each other (see Supplementary Note 6 for additional numerical simulations). On the contrary, we also plot the eigenvalues of $S$ for the case with only one unit cell (without the porous rubber sheet loading) in Fig. 3d. In this case, loss is reduced, and it does not support FP resonances to create large asymmetry in reflection amplitudes. In such a case, the eigenvalues are non-degenerate and are located near to the unit circle corresponding to the lossless case.

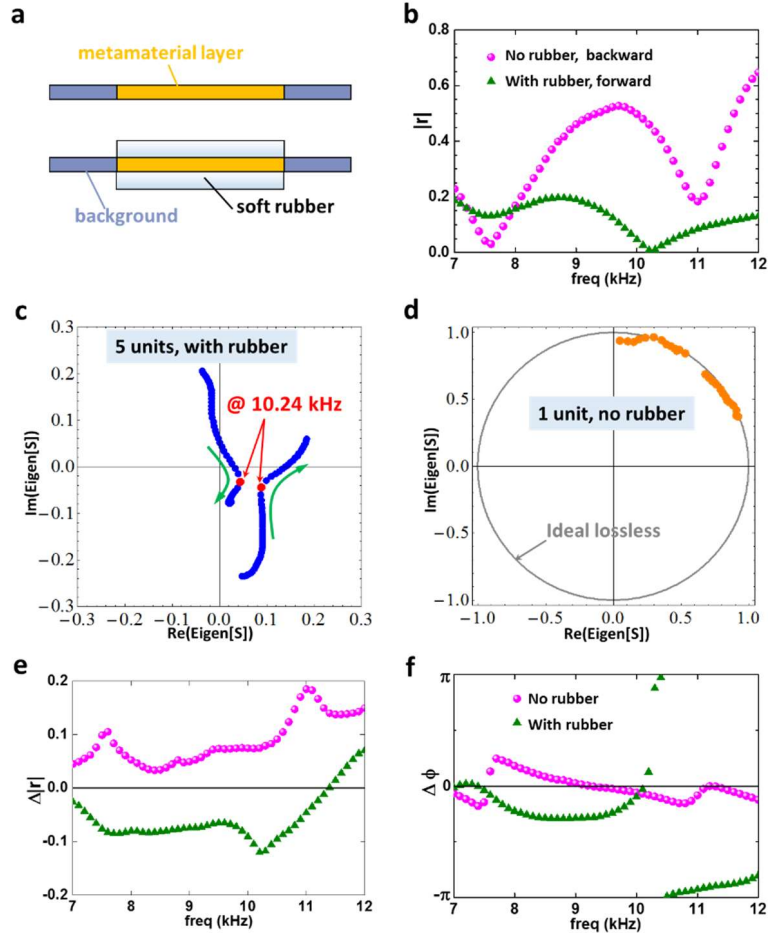

**Figure 3 | Tunable asymmetry by loading lossy porous rubber. a**, Schematic of adding loss by attaching soft porous rubber patches to the metamaterial layer. **b**, Measured amplitudes of the reflected waves $|r|$ with and without porous rubber patches. **c**, Eigenvalues of the $S$ matrix at different frequencies for a 5-unit-Willis metamaterial layer, with soft porous rubber patches attached on surfaces. Green arrows represent the trajectories of both eigenvalues along with the increase of the frequency. **d**, Eigenvalues of the $S$ matrix at different frequencies for a 1-unit-Willis metamaterial layer, no soft porous rubber pasted. The case of ideal lossless layer is a perfect unit circle. **e**, The difference of the amplitude of reflected waves $\Delta|r| = |r_f| - |r_b|$. **f**, The difference of phases for reflected waves $\Delta\phi = Arg(r_f/r_b)$ for both cases.

In fact, the loading of porous rubber sheets can be an effective way to tune not only the UZR frequency position but also the degree of reflection asymmetry, on both $\Delta|r| = |r_f| - |r_b|$ and the phase difference $\Delta\phi = \arg(r_f/r_b)$. Fig. 3e shows that measured $\Delta|r|$, which changes from positive to negative by the additional loading of porous rubber sheets. Such a change occurs over wide frequency bandwidth as well.

Fig. 3f shows the phase difference between the two directions. $\Delta\phi$ becomes large and can cover the whole $2\pi$ range. This is due to the current employment of the FP resonance in creating the asymmetry combined with the UZR (the most extreme asymmetry from an exceptional point) for the two-sheets case, in contrary to the demonstrated case airborne-acoustic bianisotropy[35], and further illustrates a non-trivial feature of our lossy Willis materials.

By extending conventional plate theory with the introduction of Willis materials, we have designed and experimentally realized effective bianisotropy in a Willis metamaterial for elastic waves using a cantilever-type resonating structure. The metamaterial induces cross-coupling between shear force and vertical displacement of the plate, or equivalently between vertical momentum and shear strain. The present design works for flexural waves on both plate-like and beam-like structures. The effective medium parameters, including the Willis coupling, are confirmed through experimental measurements of the asymmetric reflection amplitudes in the forward and backward propagation directions. Unlike previous studies on acoustic bianisotropic materials, not only asymmetric phases but also asymmetric amplitudes of reflection are observed due to the presence of loss, whose effect is further magnified through a combination of the cantilever bending resonance and a FP resonance. Interestingly, by incorporating an appropriate amount of loss, we have demonstrated the most asymmetric case in which the reflection amplitude goes to zero in one direction. Unidirectional zero reflection is achieved when approaching the exceptional point of the scattering matrix if the metamaterial is interpreted as a non-Hermitian system.

Together with the demonstrated tunability of the response (asymmetric reflection amplitudes, phases and frequency dips), the proposed metamaterials operating at the exceptional point can be applied to realize high-Q sensors[51,52], tunable metasurfaces and asymmetric wavefront control[53]. Elastic bianisotropy for scalar flexural waves can be generalized to Lamb waves for intermodal conversion control or to obtain high transmission efficiency[54-56]. The demonstration of Willis metamaterials can also be used in transformation elastic devices, as for its original motivation, to construct an elastic wave cloak based on form-invariant transformations of the elastic wave equations.

**Methods**

**Sample Fabrication and experimental setup.** All samples (structured beams with length, width and thickness 280.0, 5.0 and 2.0 $mm$, respectively) are directly printed without any assembling using a Stratasys Objet30 Pro 3D printer, based on photo-polymerization. OBJET VEROBLUE RGD840 (acrylic) is chosen as the printing material, with measured materials parameters young's modulus $E_0 = 2.95\,(1 - 0.04i)\,GPa$, Possion's ration $v_0 = 0.29$ and density $\rho_0 = 1190\,kg/m^3$ (see Supplementary Note 7). Blue tack is attached on both ends of each beam as absorbers to minimize reflected waves. Then 12mm piezo elements sounder patches are bonded at each side to act as the forward and backward sources, respectively. A pulse signal is defined then amplified by a power amplifier (KH 7602M) to generate plane waves. A Polytec PSV-400 Laser vibrometer is used to measure the wave field in the beam (160 mm length in the central area, except for the metamaterial layer), with a spatial resolution of 2.0 mm. The sampling frequency in the time domain is set at 512 kHz, and an ensemble average of 10 times is used at every scanning point. Time-domain plane wave field is tested to make sure the generation of plane waves by the piezo patches. After the fast Fourier transform (FFT), the out-of-plane wave field at each frequency is observed, which is used to further calculate the transmission and reflection spectrum (the scattering matrix $S$).

**Calculation of scattering coefficients.** Once the frequency-domain wave field is tested, we first suppose that the wave pattern can be described as $w_1 = \alpha \cdot e^{ikx} + \beta \cdot e^{-ikx}$ and $w_2 = \gamma \cdot e^{ikx} + \delta \cdot e^{-ikx}$ on the incident side and transmitted side, respectively. Here, the wave is supposed to propagate along the $x$ direction, and the origin is set at the center of the metamaterial layer. Then the complex coefficients $\alpha, \beta, \gamma$ and $\delta$ can be calculated by data fitting of the wave pattern at each frequency, for both the forward and backward case. Thus the scattering matrix is determined by

$$S = \begin{bmatrix} t_f & r_b \\ r_f & t_b \end{bmatrix} = \begin{bmatrix} \gamma_f & \gamma_b \\ \beta_f & \beta_b \end{bmatrix} \cdot \begin{bmatrix} \alpha_f & \alpha_b \\ \delta_f & \delta_b \end{bmatrix}^{-1}$$

where subscripts "$f$" and "$b$" correspond to forward and backward incident waves, respectively.

**Simulation.** All full-wave simulations performed in the paper are obtained using COMSOL Multiphysics. To compute the scattering matrix $S$, perfect matched layers (PML) are used on both end of the beam, and plane waves are generated in the forward and backward directions respectively to generate the field pattern. Then similar processes as experiments are used to calculate the scattering matrix. It is noted that the complex coefficient $D$ is zero due to the perfect absorption of waves by the PML. Therefore, we have $t_{f(b)} = \frac{\gamma_{f(b)}}{\alpha_{f(b)}}$ and $t_{f(b)} = \frac{\beta_{f(b)}}{\alpha_{f(b)}}$